\documentclass[prl,superscriptaddress,floatfix,amsmath,notitlepage]{revtex4-1}
\usepackage{graphicx}
\usepackage{color}
\usepackage{siunitx}
\usepackage[normal]{subfigure}
\usepackage[usenames,dvipsnames]{xcolor}
\usepackage{amssymb}
\newcommand{\be}{\begin{equation}}
\newcommand{\ee}{\end{equation}}

\usepackage[colorlinks=true,urlcolor=blue,linkcolor=blue,citecolor=blue]{hyperref}

\begin{document}
\title{Engineering chiral light--matter interaction in photonic crystal waveguides with slow light}

\author{Sahand Mahmoodian}
\author{Kasper Prindal-Nielsen}
\author{Immo S\"{o}llner}
\author{S{\o}ren Stobbe}
\author{Peter Lodahl}
\affiliation{Niels Bohr Institute, University of Copenhagen, Blegdamsvej 17, DK-2100 Copenhagen, Denmark}

\date{\today}


\begin{abstract}
We design photonic crystal waveguides with efficient chiral light--matter interfaces that can be integrated with solid-state quantum emitters. By using glide-plane-symmetric waveguides, we show that chiral light-matter interaction can exist even in the presence of slow light with slow-down factors of up to $100$ and therefore the light--matter interaction exhibits both strong Purcell enhancement and chirality. This allows for near-unity directional $\beta$-factors for a range of emitter positions and frequencies. Additionally, we design an efficient mode adapter to couple light from a standard nanobeam waveguide to the glide-plane symmetric photonic crystal waveguide. Our work sets the stage for performing future experiments on a solid-state platform.
\end{abstract}

\maketitle

\section{Introduction}

Waveguide quantum electrodynamics (WQED) is an attractive platform for performing experiments where a one-dimensional continuum of radiation modes  interacts strongly with quantum emitters \cite{Shen2005OL}. This platform has been used to demonstrate efficient single-photon generation \cite{Arcari2014PRL, Schwagmann2011APL}, few-photon nonlinearities \cite{Javadi2015NCOM}, and atom--atom interactions near a photonic band edge \cite{Hood2016arXiv}. The dynamics of a WQED system is ideally governed by the decay rate of the quantum emitter into the right $\Gamma_R$ and left $\Gamma_L$ propagating modes of the waveguide, however the quantum emitter generally also couples to a continuum of radiation modes outside the waveguide with a rate $\gamma$. This acts as a loss which spoils the ideal interaction of the emitter with the guided one-dimensional reservoir. Accordingly, much research has been devoted to designing light--matter interfaces that maximize  the fraction of light emitted from the quantum emitter to the waveguide mode \cite{MangaRao2007PRB, Bleuse2011PRL, LeCamp2007PRL, Lodahl2015RMP}. This is known as the beta factor $\beta = (\Gamma_R+\Gamma_L)/(\Gamma_R+\Gamma_L +\gamma)$. Nanophotonic waveguides composed of high refractive index materials have strongly confined optical modes and when patterned with photonic nanostructures can both suppress coupling to radiation modes using a photonic bandgap and further enhance coupling to the waveguide mode by reducing its group velocity $v_g$. Such systems can be efficiently combined with solid-state quantum emitters such as quantum dots \cite{Lodahl2015RMP} with near-unity values of $\beta$ being demonstrated recently \cite{Arcari2014PRL}.

An exciting development in WQED systems is the demonstration of chiral light--matter interaction, i.e., when the symmetry between emission into the right- and left-propagating modes is broken $\Gamma_R \neq \Gamma_L$ \cite{Petersen2014Science, Sollner2015NNANO, Coles2016NCOM, Lodahl2016arXiv}. Since the Hamiltonian in these systems is not symmetric under time-reversal \cite{Stannigel2012NJP}, they can be used to construct optical isolators and circulators \cite{Xia2014PRA, Sollner2015NNANO, Sayrin2015PRX, Scheucher2016arXiv} as well as to dissipatively prepare entangled states in a cascaded spin network \cite{Stannigel2012NJP, PichlerPRA2015}. In the extreme case this interaction can also become unidirectional, i.e., the emitter couples only to a single direction. Unidirectional emission occurs when the transition dipole moment of the emitter couples to the waveguide mode in one direction while being orthogonal to the mode in the other direction. This has been achieved by engineering the modes of the nanophotonic waveguide to have an in-plane circular polarization. The modes of a dielectric waveguide are time-reversal symmetric and therefore $\omega_{-k} = \omega_k$ and $\mathbf E_{-k} (\mathbf r) = \mathbf E_{k}^* (\mathbf r)$ implying that counter-propagating modes are counter circulating. An ideal WQED system with a chiral light-matter interface exhibits unidirectional emission and also minimizes the coupling fraction to radiation modes. The fraction of light coupled to a single directional mode is characterized by the directional beta factor $\beta_{R/L} = \Gamma_{R/L}/(\Gamma_R+\Gamma_L+\gamma)$ \cite{Lodahl2016arXiv}, where the subscripts $R$ and $L$ refer to right or left respectively. In this manuscript we show that an ideal chiral light-matter interface can be achieved in a glide-plane waveguide. Unlike in previous work \cite{Sollner2015NNANO, Young2015PRL, Lang2016arXiv}, we show that in-plane circular polarization can be preserved in photonic crystal waveguides with glide-plane symmetry (GPWs) for group velocities as low as $v_g \sim c/100$ and that the waveguide has only a single mode. This requires carefully engineering the glide-plane waveguide to ensure that it only has a single mode while preserving its symmetry. By computing the emission properties of a dipole embedded in the waveguide, we show that the decay rate of emission to a single directional mode can be enhanced by a factor of 5 while maintaining directional beta factors of $\beta_{R/L}>0.99$. Finally, we design an efficient mode adapter to couple between a regular suspended nanobeam waveguide and our GPW.

\section{Engineering light-matter interaction in glide-plane waveguides}

\begin{figure}[!t]
\centering
\includegraphics[width=0.8\columnwidth]{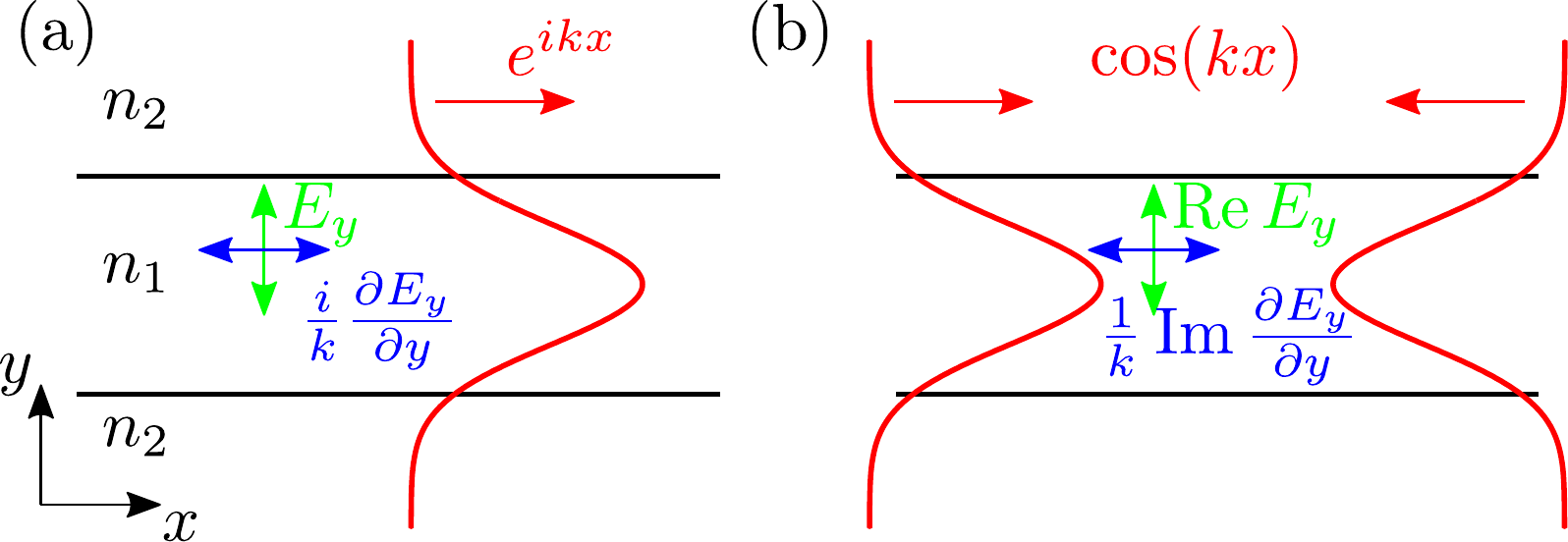}
\caption{\label{fig:designPrinciples} Schematics of travelling and standing waves in a two-dimensional nanophotonic waveguide. (a) A TE mode (electric field in $x$-$y$ plane) of a high-index waveguide ($n_1>n_2$) propagating to the right with propagation constant $k$. This mode has a longitudinal electric field component that is $\pi/2$ out of phase with the transverse component and its magnitude is related to the confinement of the mode. With appropriate design the field can be circularly polarized. (b) An interference pattern composed of counter-propagating modes of the waveguide in (a) with wavevectors $\pm k$. The resultant mode has electric fields that are real and the cannot have in-plane circular polarization.}
\end{figure}

Before presenting our design we briefly discuss the requirements for realizing a waveguide whose modes have an in-plane circular polarization. The first requirement is that the waveguide modes have a longitudinal field component thus requiring non-paraxial wave propagation. The longitudinal field component must also be $\pi/2$ out of phase with the transverse field component. Fortunately, it has been shown that the magnitude of the longitudinal component of the electric field of a confined mode is proportional to the strength of its transverse confinement \cite{Bliokh2012PRA, Lodahl2016arXiv} and that Gauss's Law ensures that the longitudinal component is $\pi/2$ out of phase with the transverse component. This is illustrated in a simple example in Fig.~\ref{fig:designPrinciples}(a) which shows a schematic of the field in a two-dimensional nanophotonic waveguide. Since the geometry is piecewise homogeneous, within the waveguide Gauss's Law requires that $\frac{\partial E_x}{\partial x}+\frac{\partial E_y}{\partial y}=0$. The field of the waveguide is given by $\mathbf E_{k} (\mathbf r) = \left[E_x(y) \hat{\mathbf x} + E_y(y) \hat{\mathbf y} \right]e^{i k x}$ and therefore the waveguide mode must possess a longitudinal field component given by $E_x = \frac i k \frac{\partial E_y}{\partial y}$. We note that this argument only holds if eigenstates of the waveguide are forward and backward propagating modes. If the forward and backward propagating states are coupled, for example by creating a cavity, the mode becomes a linear combination of these states. The limit of where the eigenmodes are complete standing-waves is shown in Fig.~\ref{fig:designPrinciples}(b). Here, since the mode is a linear combination of the forward $\mathbf E_{k}$ and backward $\mathbf E_{-k} = \mathbf E_{k}^*$ propagating modes, it becomes a real-valued field and cannot possess an in-plane circular polarization. 

Although photonic-crystal waveguide (PCW) modes are more complex than the simple schematic shown in Fig.~\ref{fig:designPrinciples}, the same arguments can be used to determine whether its modes have an in-plane circular polarization. Importantly, the presence of a band-edge occurring at the Brillouin zone (BZ) boundary $k=\pi/a$, where $k$ is the Bloch wavevector and $a$ is the lattice period, sets the same restrictions on the mode profile \cite{Sukhorukov2009JoptA}. In particular, if the dispersion is quadratic the field becomes a real-valued standing wave and does not have circularly polarized modes \cite{Lang2016arXiv}. This follows due to time-reversal symmetry and crystal periodicity giving the relation
\begin{equation} \label{eq:bandEdge}
\mathbf E_{\pi/a}^*(\mathbf r) = \mathbf E_{-\pi/a}(\mathbf r) = \mathbf E_{\pi/a}(\mathbf r),
\end{equation}
as the two zone edges are separated by a reciprocal lattice vector $2\pi /a$. This means that regular PCWs have modes that are standing waves near the band-edge and cannot have an in-plane circular polarization. We emphasize that the behaviour near the band edge is of importance as this is the regime of slow light.

\begin{figure}[!t]
\centering
\includegraphics[width=\columnwidth]{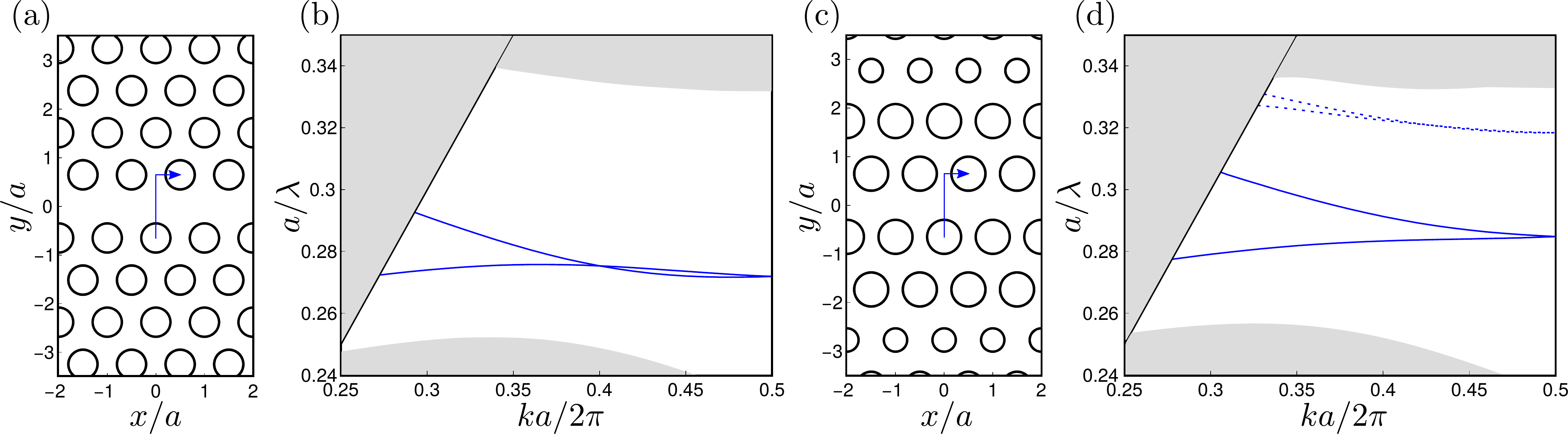}
\caption{\label{fig:geomDisp} Glide-plane waveguide geometries and their dispersion curves. (a) A GPW in a triangular lattice with hole-to-hole-centre width of $0.75 \sqrt{3} a$ and (b) its dispersion curve. The blue lines are the waveguide modes and the grey shading indicates modes not guided by the waveguide. (c) A GPW dispersion engineered to have only a  single mode propagating in each direction and (d) its dispersion curve. The dashed blue lines show other modes that are not of interest. See main text for waveguide parameters.}
\end{figure}

This constraint can be overcome by introducing glide-plane symmetry to the waveguide. A glide-plane operation is a reflection about a plane followed by a translation. Figure \ref{fig:geomDisp}(a) shows a GPW that is invariant under a reflection about the $x$-$z$ plane followed by a translation of $a/2$ along the $x$-direction. We denote the glide-plane operator $\hat G = \hat{T}_{x=a/2} \hat{R}_y$, where $\hat{T}_{x=a/2}$ is the half-period translation operator and $\hat{R}_y$ is the reflection operator. Since the GPW is invariant under the operation $\hat G$, its electric field eigenstates $\mathbf E_k (\mathbf r)$ are also eigenstates of $\hat G$. To find the eigenvalues of $\hat G$, we note that two glide-plane operations correspond to a translation along a unit cell. Therefore $\hat G^2 \mathbf E_k(\mathbf r) = e^{i k a} \mathbf E_k(\mathbf r)$ and thus $\hat G \mathbf E_k(\mathbf r) = \pm e^{i k a/2} \mathbf E_k(\mathbf r)$. Importantly, at the Brillouin zone edge $k=\pi/a$ and thus the eigenvalues of the glide-plane operator are $\pm i$. This means that two positions in the GPW separated by a glide-plane symmetry operation differ by a factor of $\pm i$ and therefore the field cannot be entirely real at the zone edge. To satisfy time-reversal symmetry there must be a second mode at the zone edge with the same frequency which is related to the first through
\begin{equation} \label{eq:GPWrelation}
\mathbf E_{1,\pi/a} (\mathbf r) = \mathbf E_{2,\pi/a}^* (\mathbf r).
\end{equation}
In contrast to Eq.~\ref{eq:bandEdge} this generally implies that $\mathbf E_{1,\pi/a} (\mathbf r)$ and $\mathbf E_{2,\pi/a} (\mathbf r)$ are complex fields and does not constrain their polarization properties. Under these conditions the modes generally have a linear dispersion relation with equal and opposite group velocities at the Brillouin zone boundary.

A schematic of a GPW with a hole-to-hole-centre width of $0.75 \sqrt{3} a$ between the centre of holes on the two sides of the waveguide is shown in Fig.~\ref{fig:geomDisp}(a) with its dispersion curve shown in Fig.~\ref{fig:geomDisp}(b). For all calculations the radii of the cylinders are $r=0.3a$ unless otherwise stated, the membrane thickness is $t=0.6154 a$ with a refractive index of $n=3.464$ corresponding to GaAs at cryogenic temperatures. All dispersion curves are for three-dimensional structures and are computed using freely available software \cite{Johnson2001OE}.  As predicted, the combination of time-reversal symmetry and glide-plane symmetry causes a degeneracy at the BZ boundary. Unfortunately, the dispersion curve has multiple modes as the two curves cross in Fig.~\ref{fig:geomDisp}(b). We note the difference to topologically-protected edge states, which can be designed to guarantee a single Dirac point \cite{Hasan2010RMP}. This is problematic because, due to Eq.~\ref{eq:GPWrelation}, the two modes will tend to have opposite handed circular polarizations at a given position and therefore at such a position an emitter will couple to two counter-propagating modes destroying the directionality of the system. 

To make this waveguide optimum for chiral light-matter interaction we must engineer the dispersion of the waveguide \cite{Frandsen2005OE} to ensure that the waveguide bands are single moded. This involves changing the radius and position of the holes around the waveguide. Since the electric field $\mathbf E_k (\mathbf r)$ has different field profiles along the dispersion curve, perturbations to the holes affect the frequency of the modes at different wavevectors differently, i.e., the perturbation matrix element $\langle \mathbf E_k | \delta \epsilon | \mathbf E_k \rangle$ varies with $k$, where $\delta \epsilon$ is some perturbation to the permittivity distribution of the GPW. We vary the hole positions and radii until the bands have the desired single mode dispersion. The GPW schematic shown in Fig.~\ref{fig:geomDisp}(c) has been engineered to have the single mode dispersion curve shown in Fig.~\ref{fig:geomDisp}(d). We end up with a design where rows two to four are shifted outwards by $l_2 = 0.25 a \sqrt{3}/2$, $l_3 = 0.2 a \sqrt{3}/2$, and $l_4 = 0.1 a \sqrt{3}/2$, and the radii of the first three rows of holes are modified to $r_1 = 0.35 a$, $r_2 =0.35 a$, and $r_3 = 0.24 a$.

\section{Results}

\begin{figure}[!t]
\centering
\includegraphics[width=\columnwidth]{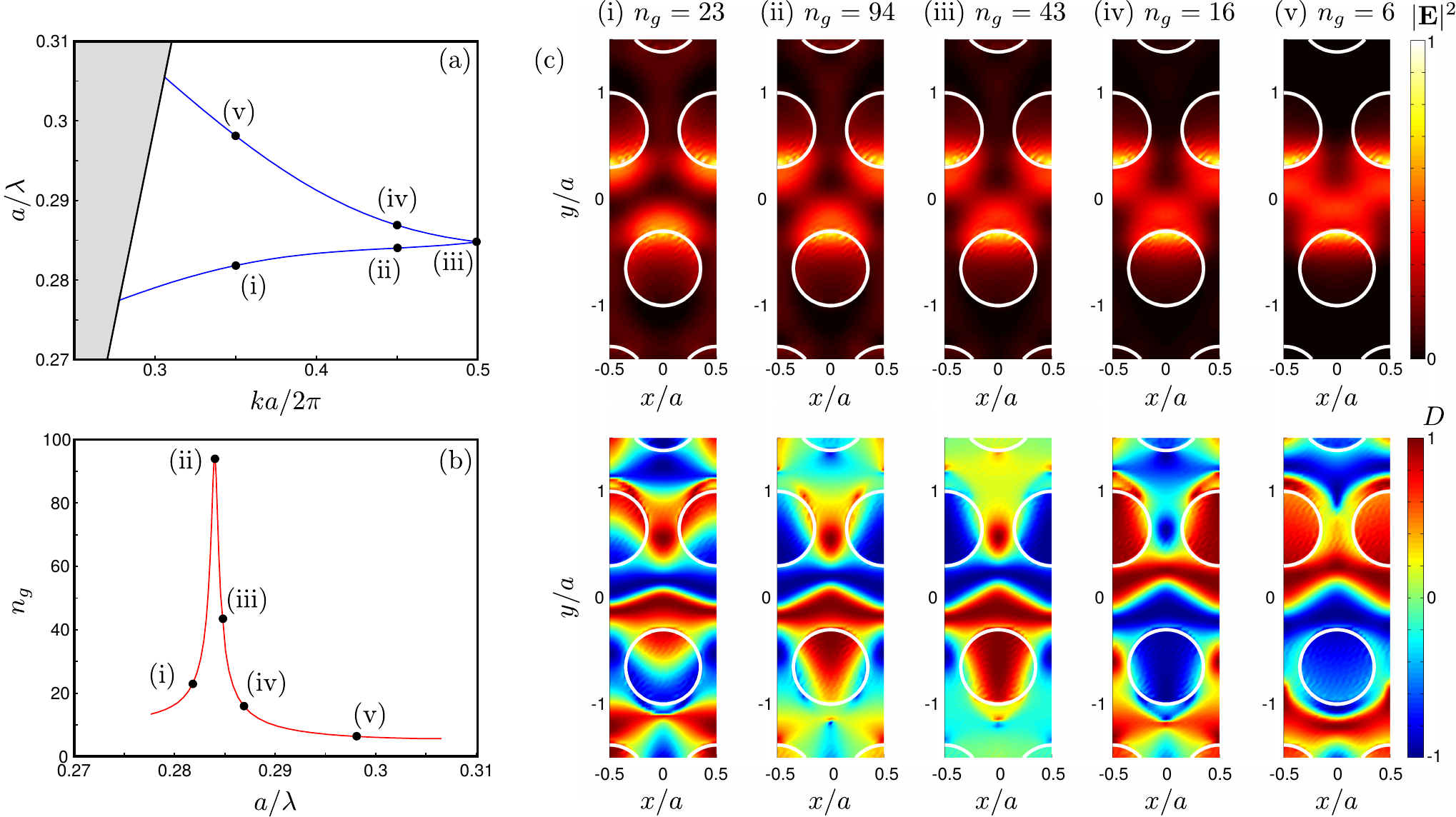}
\caption{\label{fig:resultsFields} (a) The dispersion curve of the GPW and (b) the magnitude of the group index of its modes versus frequency for the structure in Fig.~\ref{fig:geomDisp}(c). (c) The electric field intensity $|\mathbf E|^2$ (top row) and the directionality factor $D$ (bottom row) for modes along the dispersion curve with normalized frequencies and group indices (i) $a/\lambda=0.282$ and $n_g = 23$ (ii) $a/\lambda = 0.284$ and $n_g=94$ (iii) $a/\lambda=0.285$ and $n_g=43$ (iv) $a/\lambda = 0.287$ and $n_g=16$ (v) $a/\lambda = 0.298$ and $n_g=6$.}
\end{figure}

Figure~\ref{fig:resultsFields}(a) shows a close-up of the dispersion curve of the GPW and Fig.~\ref{fig:resultsFields}(b) shows the slow-down factor given by the group index $n_g= c/v_g$ where $c$ is the speed of light in vacuum. The group index becomes as large as $n_g=94$ implying a significant increase in the light-matter interaction. The electric field intensity of the modes $|\mathbf E(\mathbf r)|^2$ is shown in the top panel of Fig.~\ref{fig:resultsFields}(c) while the bottom panel shows the directionality  $D = \left[|\mathbf E^* (\mathbf r) \cdot \hat{ \mathbf l}|^2 - |\mathbf E^* (\mathbf r) \cdot \hat{ \mathbf r}|^2\right]/|\mathbf E (\mathbf r)|^2$, with $\hat{\mathbf r} = (\hat{\mathbf x}+ i \hat{\mathbf y})/\sqrt{2}$ and $\hat{\mathbf l} = (\hat{\mathbf x} - i \hat{\mathbf y})/\sqrt{2}$ where $\hat{\mathbf x}$ and $\hat{\mathbf y}$ are unit vectors. Since there is only one GPW mode, $D$ determines the fraction of light emitted to the right or left within the waveguide, i.e., $D=1$ implies the field is entirely left-hand circularly polarized,  $D=-1$ implies the field is entirely right-hand circularly polarized while $D=0$ corresponds to linear polarization. Therefore an emitter with a circularly polarized transition dipole moment positioned at a point with $|D|=1$ will emit unidirectionally within the waveguide mode. We highlight that there are many positions with $| D|\sim 1$ for all frequencies and even for group indices up to $n_g=94$. Importantly the shape of the Bloch modes is also such that the field has a significant fraction of its maximum intensity near the positions where the field is circularly polarized. A quantum dot that is well-coupled to the mode is therefore also likely to be at a position of high directionality.


\begin{figure}[!t]
\centering
\includegraphics[width=0.9\columnwidth]{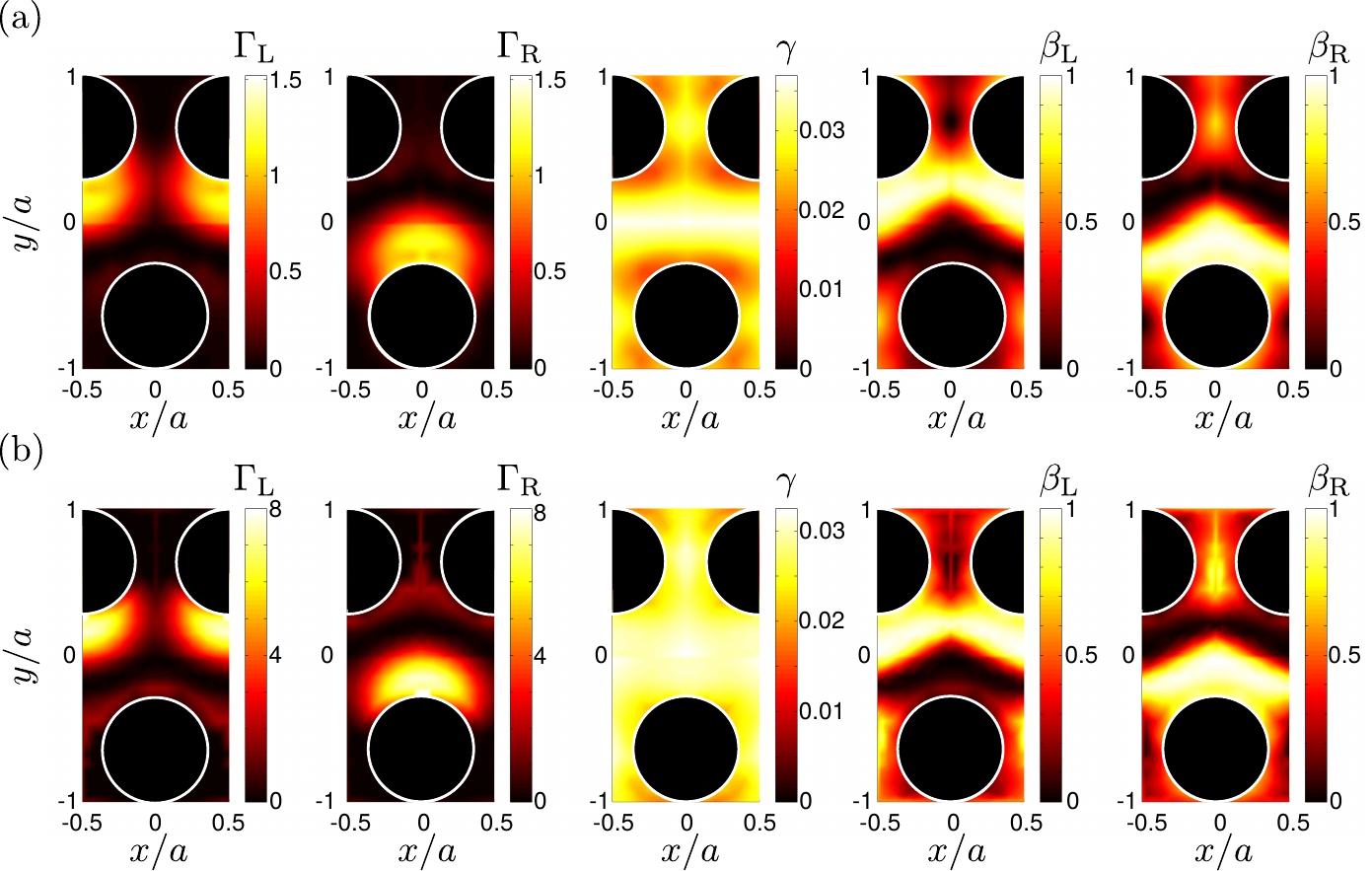}
\caption{\label{fig:resultsDipole} Radiation properties of a left-hand circularly polarized dipole in the GPW as a function of position for (a) $a/\lambda =0.29$ corresponding to $n_g = 10$ and (b) $a/\lambda = 0.284$ corresponding to $n_g=94$. The panels show (from left to right) the relative decay rate to the left-propagating mode $\Gamma_L$, the right-propagating mode $\Gamma_R$, the unguided radiation reservoir $\gamma$, the $\beta$-factor of the left propagating mode $\beta_L$, and the $\beta$-factor of the right propagating mode $\beta_R$. All decay rates are normalized to emission in a homogeneous medium with refractive index $n=3.4638$.}
\end{figure}

To fully appreciate the properties of the GPW for chiral light-matter interaction we quantify the fraction of emission that couples to the guided modes of the waveguide and the emission that couples to other modes. Figure \ref{fig:resultsDipole} shows the results of computing the emission profile of a dipole for different positions within a unit cell of the GPW for two different frequencies. These computations were carried out using frequency-domain finite-element modelling software. Further details of these calculations can be found in \cite{Javadi2014InPrep}. Unlike the eigenmode calculations shown in Fig.~\ref{fig:resultsFields}, these calculations enable us to also quantify the coupling to the radiation reservoir and extract the decay rate $\gamma$. Since these computations are considerably more time consuming, and because quantum dots cannot lie within the air holes, we have restricted the computation to positions within the GaAs.  Both at $n_g=10$ and at $n_g=94$, corresponding to Figs.~\ref{fig:resultsDipole}(a) and (b) respectively, the directional beta factors approach unity. This is because the large refractive index contrast and the photonic bandgap inhibit coupling to the unguided radiation reservoir and the emission is dominated by the waveguide mode. We find that at $n_g=94$ the directional coupling rates $\Gamma_R$ and $\Gamma_L$ exhibit strong Purcell enhancement of up to 8. Purcell enhancement can help overcome decoherence processes in solid-state emitters \cite{Lodahl2015RMP}. At this frequency we have found positions where $\beta_{R/L}>0.99$ while $\Gamma_{R/L}>5$. Such parameters indicate that the GPW constitutes an almost ideal WQED geometry with chiral light-matter interaction. 

We note that chiral light-matter interaction in solid-state systems has been demonstrated in regular nanobeam waveguides \cite{Coles2016NCOM}. By computing the emission properties in this waveguide, we have found that  the directionality factor approaches unity, but the $\beta$ factor has a maximum of $\beta_{R/L}\sim 0.7$ (see Appendix A).

\section{An efficient mode adapter}

\begin{figure}[!t]
\centering
\includegraphics[width=\columnwidth]{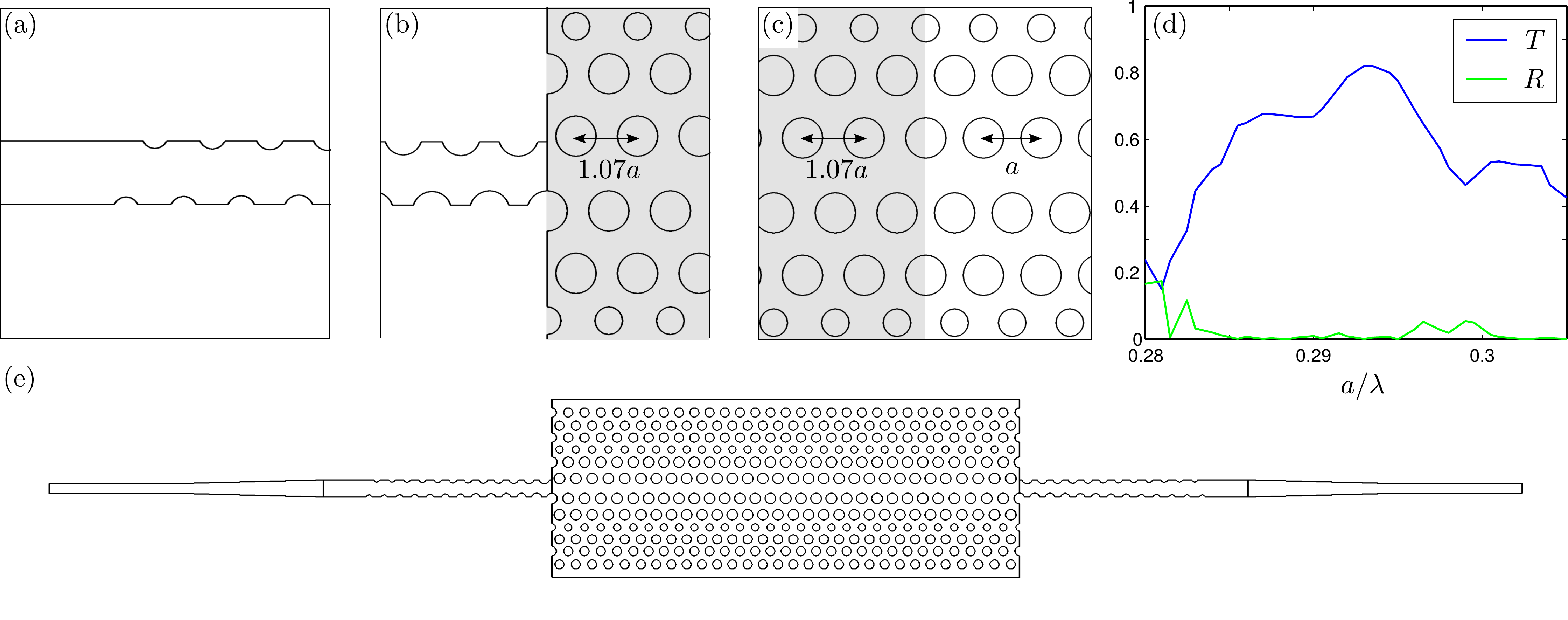}
\caption{\label{fig:transFig} The mode adapter and its transmission spectrum. The mode adapted is composed of three parts: (a) circular indents are introduced adiabatically. to the nanobeam waveguide to break the reflection symmetry. (b) The glide-plane symmetric nanobeam waveguide is coupled into a GPW whose period is $1.07$ times the period of the regular GPW. See main text for detailed parameters. (c) The stretched GPW is coupled directly to the GPW. (d) Normalized transmitted and reflected powers through the GPW with $1-T-R$ corresponding to power lost through scattering to other modes. (e) Computation domain of the GPW. The GPW contains 20 periods with period $a$ with 5 periods with the stretched lattice $1.07a$ on both sides.}
\end{figure}

For most quantum photonics applications, it is necessary to couple light off chip \cite{Mahmoodian2016arXiv}. Coupling light efficiently off chip to a fiber requires mode-matching of the tightly confined modes of the on-chip waveguides to a much larger mode confined in the core of a single-mode fiber. This is usually achieved by using inverse tapers \cite{Cohen2013OE} or specially designed gratings \cite{Taillaert2006JJAP}. These tapers and gratings usually complemented with standard ridge waveguides or underetched nanobeam waveguides, therefore, in order to couple light propagating in a GPW off-chip we must first demonstrate efficient coupling to a nanobeam waveguide. Unfortunately, because of the GPW's symmetry, its modes cannot be coupled to a nanobeam waveguide directly and a mode adapter is required.

Figure \ref{fig:transFig}(a)-(c) shows three segments of a mode adapter that couples light from a suspended nanobeam waveguide of width $w=a$ to a GPW. The purpose of the first segment in Fig.~\ref{fig:transFig}(a) is to adiabatically break the reflection symmetry of the nanobeam about the $x$-$z$ plane but preserve glide-plane symmetry in the nanobeam. We do this by introducing cylindrical indentations into the side of the nanobeam. These are made by removing the high index material of the nanobeam that overlaps with circles positioned at $\pm 0.75a \sqrt{3}/2$  whose radius starts at $R_1 = 0.23a$ and increases linearly until the twelfth cylinder which has radius $R_{12} = 0.35 a$ which is the same as the radius of the first row of holes in the GPW. After this section of the taper the mode has the same symmetry as the GPW. Figure \ref{fig:transFig}(b) shows the second part of the mode adapter where the glide-plane symmetric nanobeam couples to a GPW whose lattice constant has been stretched to $1.07a$. The lattice stretching has been used for regular photonic crystal waveguides \cite{Hugonin2007OL} and has the effect of shifting the dispersion curve lower. For the frequencies of interest this means that the group index of the stretched region is lower and ensures there is not a significant impedance mismatch between the nanobeam and the stretched GPW. This can then be directly coupled to the GPW with lattice constant $a$. Just as in regular PCWs, evanescent modes help match the mode of the stretched GPW to the regular GPW to ensure efficient coupling \cite{deSterke2009OE}. The transmitted power through the  mode adapter to a GPW and out through a mode adapter again is shown in Fig.~\ref{fig:transFig}(d). We emphasize that the light propagates through the mode adapter twice. The fraction of transmitted power is comparable to coupling from a nanobeam waveguide to a regular PCW and peaks at 0.82. Here the losses are dominated by scattering into other modes. The transmission through the structure exceeds 0.6 between $a/\lambda = 0.285$ and $a/\lambda=0.297$. 

\section{Conclusion}
Future experiments exploiting chiral light-matter interaction such as dissipatively generating entangled dark states \cite{PichlerPRA2015}, performing two-qubit parity measurements \cite{Mahmoodian2016arXiv}, or realizing non-reciprocal photon transport \cite{Sollner2015NNANO} all require ideal chiral interactions between a one-dimensional radiation bath and quantum emitters. We have shown that glide-plane waveguides provide unidirectional emission and near-unity $\beta$-factors and that, when interfaced with high-quality solid-state quantum emitters, can form a platform for performing the next generation of experiments.

\section*{Funding}
We gratefully acknowledge financial support from the Villum Foundation, the Carlsberg Foundation, the Danish Council for Independent Research (Natural Sciences and Technology and Production Sciences), and the European Research Council (ERC Consolidator Grants ALLQUANTUM and QIOS).

\section*{Acknowledgments}
The authors would like to thank Alisa Javadi and Leonardo Midolo for enlightening discussions.

\newpage

\appendix

\section{Appendix}

\begin{figure}[!t]
\centering
\includegraphics[width=\columnwidth]{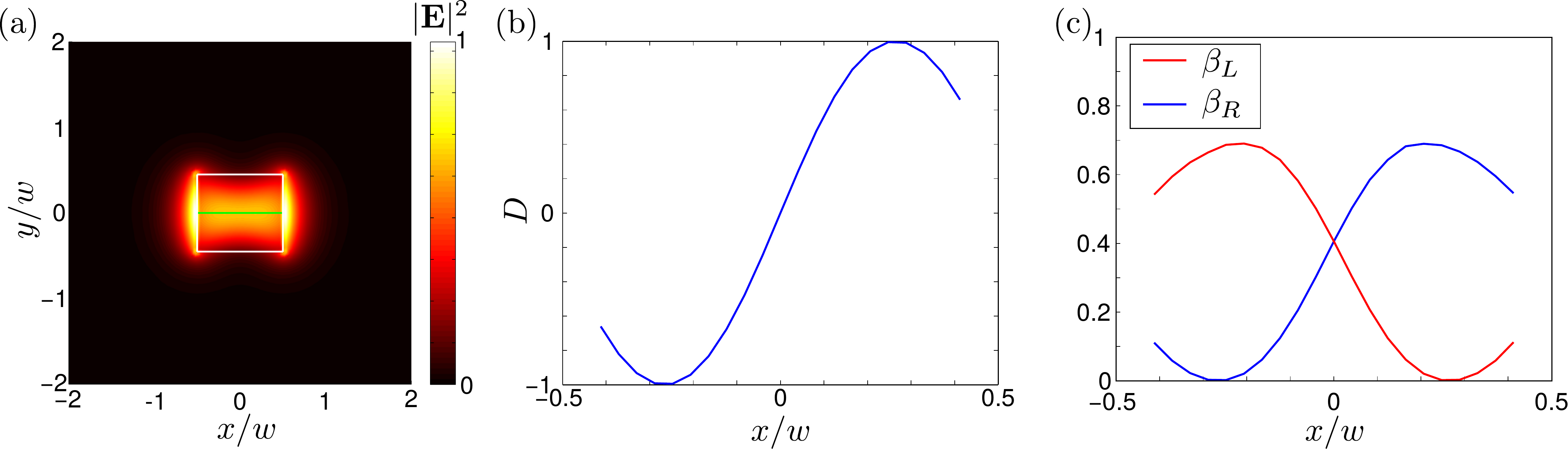}
\caption{\label{fig:nanobeam} Chiral light-matter interaction in a nanobeam waveguide. (a) Electric field intensity along the cross section of a nanobeam waveguide with width $w$ and thickness $t=0.9 w$. (b) Directionality function of a left-hand circularly polarized dipole positioned along the green line shown in figure (a). (c) Directional beta factors for a left hand circularly polarized dipole along the waveguide. All computations are at a normalized frequency of $w/\lambda = 0.215$.}
\end{figure}

Nanobeam waveguides have also been used to demonstrate chiral light-matter interaction with near-unity values of directionality being reported \cite{Coles2016NCOM}. Here we briefly evaluate their properties numerically. Figure \ref{fig:nanobeam}(a) shows the electric field intensity of the mode of a nanobeam waveguide at a frequency where the waveguide has a single mode forward and backward mode. By computing the emission properties of a dipole with positions along the green line in  Fig.~\ref{fig:nanobeam}(a) we extract the directionality $D$ and directional beta factors $\beta_{R/L}$. Figure \ref{fig:nanobeam}(b) shows that the nanobeam waveguide contains positions where the emission can be completely unidirectional, which is in agreement with the experimental results in \cite{Coles2016NCOM}. However, the absence of slow light and a photonic bandgap in the nanobeam waveguide causes a larger fraction of the emission to couple to radiation modes outside of the waveguide. This is encapsulated by the directional beta factor plotted in Fig.~\ref{fig:nanobeam}(c). Here the directional $\beta_R$ and $\beta_L$ approach values only as large $0.7$. Although the nanobeam waveguide provides a simple geometry for obtaining chiral light-matter interaction, it is less efficient than the GPW.


\bibliography{bigBib}{}
\bibliographystyle{unsrt}

\end{document}